\documentstyle[amssymb,aps,psfig]{revtex}
\newcommand{\be}{\begin{equation}}
\newcommand{\ee}{\end{equation}}
\newcommand{\br}{\begin{eqnarray}}
\newcommand{\er}{\end{eqnarray}}
\newcommand{\nn}{\nonumber}
\newcommand{\bd}{\begin{displaymath}}
\newcommand{\ed}{\end{displaymath}}
\newcommand{\ovl}{\overline}

\newcommand{\bfig}{\begin{figure}}
\newcommand{\efig}{\end{figure}}

\def\w{\omega}
\def\la{\lambda}
\def\alf{\alpha}

\def\lb#1{\label{#1}}
\def\3cdot{\cdot \cdot \cdot}
\def\rf#1{\ref{#1}}

\def\g{\gamma}

\def\half{\frac{1}{2}}
\def\l{\lambda}

\def\om0{\omega _0}
\def\Om0{\Omega _0}

\def\x{\times}
\def\rg{\rangle}
\def\ra{\rangle}
\def\lg{\langle}
\def\text#1{{\rm{#1}}}
\def\sig{\sigma}

\def\->{\rightarrow}
\def\=>{\Rightarrow}
\def\-->{\longrightarrow}
\def\==>{\Longrightarrow}

\def\rpar{\right)}
\def\lpar{\left(}
\def\lbk{\left[}
\def\rbk{\right]}
\def\lbr{\left\{}
\def\rbr{\right\}}
\def\dag{\dagger}
\def\ox{\otimes}

\def\pr{^\prime}
\def\pr2{^{\prime\prime}}
\def\rf#1{(\ref{#1})}

\def\bfig{\begin{figure}}
\def\efig{\end{figure}}

\begin{document}
\title{Controlling quantum entanglement through photocounts}
\author{M. C. de Oliveira\thanks{E-mail: marcos@df.ufscar.br}, L. F. da Silva, and S. S. Mizrahi
\thanks{E-mail: salomon@df.ufscar.br}}
\address{Departamento de F\'{\i}sica, CCET, Universidade Federal de S\~{a}o
Carlos,\\ Via Washington Luiz km 235, 13565-905, S\~ao Carlos, SP,
Brazil.}
\date{\today}
\maketitle
\begin{abstract}
We present a protocol to generate and control quantum entanglement
between the states of two subsystems (the system ${\cal S}$) by making
measurements on a third subsystem (the monitor ${\cal M}$),
interacting with ${\cal S}$. For the sake of comparison we consider
first an ideal, or instantaneous projective measurement, as postulated
by von Neumann. Then we compare it with the more realistic or
generalized measurement procedure based on photocounting on ${\cal
M}$. Further we consider that the interaction term (between ${\cal S}$
and ${\cal M}$) contains a quantum nondemolition variable of ${\cal
S}$ and discuss the possibility and limitations for reconstructing the
initial state of ${\cal S}$ from information acquired by photocounting
on ${\cal M}$.

\pacs{03.67.-a, 03.65.Bz, 42.50.-p}
\end{abstract}

%
\section{Introduction}
%
Information processing can be largely improved when quantum properties
are used for encoding both, bits and channels \cite{qinf}. While bits
are encoded in mutually orthogonal states of a quantum system, quantum
channels use the ability to set systems in entangled states.
Entanglement of states is a characteristic quantum correlation that,
in principle, can be produced in post interacting quantum systems
\cite{ent}. To use this quantum resource for information processing
one has to be able first to produce and then to control the amount of
entanglement of a finite number of quantum systems.  A fundamental
open problem in quantum information is the characterization and
classification of mixed entangled states of a multipartite systems
\cite{cirac}. Nowadays the most accessible and controllable source of
quantum entanglement has been the electromagnetic field, through
parametric down-conversion processes in non-linear crystals
\cite{kwiat,white}. Recently, internal atomic states entanglement have
also been considered in distinct experiments \cite{hagley,rauschenbeutel,sackett,nature}.

We propose a consistent scheme for generating and controlling
entangled states of two subsystems (that we call ${\cal S}$): (i) If
the subsystems do interact, their initial states should be adequately
prepared such that the interaction $V_S$ does not entangle their
states during the time evolution. (ii) A third quantum subsystem, the
monitor ${\cal M }$, is coupled (through $V_{SM}$) to ${\cal S}$;
${\cal M }$ should be the only subsystem responsible for entangling
the states of ${\cal S}$, thus formally, $[V_S,V_{SM}]=0$ is a
necessary condition. Then follows an operational procedure or
protocol: (iii) First one choose an observable ${\cal K}$ of ${\cal
M}$ and after an elapsed time $t$, from the beginning of the
interaction (between ${\cal S}$ and ${\cal M}$), it is measured and
the interaction is turned off; the eigenvalue outcome determines the
entangled state in which ${\cal S}$ is left.

In an ideal projective measurement the measured eigenvalues $k$ occur
with a certain probability, so the entangled state of ${\cal S}$
obtained through reduction, cannot, in principle, be chosen. However,
if the experimentalist is able to control the outcomes of ${\cal M}$
to be read, then, controlling the degree of entanglement in ${\cal S}$
becomes possible and the protocol becomes feasible. Interestingly, in
a realistic measurement, when quanta are counted, it is possible to
control the outcome of the monitor: One turns on the interaction
between ${\cal S}$ and ${\cal M}$ and when $k$ quanta are counted then
one knows in which state the system ${\cal S}$ is left. However, the
necessary interval of time for counting $k$ quanta is probabilistic:
if the experiment is repeated, the same number of quanta may be
registered within another time interval. Thus, for reproducing the same
state one should repeat the experiment such that the counted photons be
the same within the same time interval. The present status for
generating entangled states is very different from this protocol, the
experiments  are based on projective measurements \cite{kwiat,white}.

The realization of the proposed scheme and protocol are considered in
the following physical system: the subsystems constituting ${\cal S} $
may consist of two interacting (but not necessarily) electromagnetic
(EM) fields, modes A and B, coupled to the monitor ${\cal M}$, a third
EM field, the mode C. The entanglement in ${\cal S}$ is created by a
continuous destructive photocount on C and the control is fulfilled by
turning off the interaction when a certain predetermined number $k$ of
photons become registered, thus the system ${\cal S}$ is left in an
entangled state which is essentially characterized by $k$. This
proposal is detailed in the following sections: In Sec. II we present
our model and write the Hamiltonian for the system ${\cal S}$, the
monitor ${\cal M}$ and their interaction, we also give the
time-dependent state vector of the whole system. In Sec. III we
describe the measurement process in two different approaches for sake
of comparison, the ideal and the realistic: (i) The ideal or
instantaneous projective measurement on field C, assumes the
statevector reduction by projection as postulated by von Neumann,
leading to an entangled (regarding the fields A and B) pure states.
(ii) Then, more realistically, we consider the measurement as a
sequential photocount process, on field C, as proposed by Srinavas and
Davies \cite{davies}. This leads to an entangled density operator for
fields A and B. We show that only the second approach allows
controlling the degree of entanglement of the AB fields, depending on
the detector counting rate ($\g$), the number of counted photons ($k$)
and counting time. In Sec. IV we show that due to the nondemolishing
character of the coupling between ${\cal S}$ and ${\cal M}$, the total
photon number $N$ of ${\cal S}$ can be inferred (without altering $N$)
by averaging over counted photons of mode C. We also analyze the
information one gets about the initial state of ${\cal S}$ from the
counting process on ${\cal M}$; examples are presented and discussed.
In Sec. 5 we present a summary and conclusions.
%
\section{Model}
%
The problem of production and control of state entanglement between two
subsystems is based on the physical paradigm of four interacting EM
fields. Couplings of EM modes are made possible in nonlinear media and
phenomena such as parametric down and up conversion appearing when a
response of second order nonlinearity in crystal polarization is
present and four-wave mixing (third order nonlinearity) occurs in a
Kerr medium \cite{yariv}. Also, recently a strong field-field
interaction of few photons, induced by non-resonant interactions
between fields and a Cs atom, was observed experimentally
\cite{kimble}. This observation led to a proposal for attaining
high-nonlinearities with single atoms \cite{parkins}. The dynamics of
the fields here considered consists of two processes: (i) a second
order nonlinear process coupling the modes A and B (${\cal S}$),
assisted by a classical pump field \cite{milburnli,scully} and (ii) a
four-wave mixing, with modes, A, B and C (treated as quantized fields)
coupled to a fourth classical intense field. The system dynamics is
described by the Hamiltonian
\begin{eqnarray}\label{1}
 H &=&\hbar \w_a a^\dagger a+\hbar\w_b b^\dagger b
+\hbar\w_cc^\dagger c+\hbar\la \left(a^\dagger b e^{i\nu t}+ab^\dagger
e^{-i\nu t}\right)\nonumber\\ &&+ \hbar\chi (a^\dagger a+b^\dagger
b)(c e^{-i\nu ' t}+c^\dagger e^{i\nu ' t})\, . \lb{ham1}
\end{eqnarray}
The total number of photons of modes A and B, $\hat{N} \equiv
\hat{n}_a + \hat{n}_b = a^\dagger a+ b^\dagger b$ is a quantum
nondemolition (QND) variable. This is a quite important feature
because while  measuring (destructively) the mode C, an inference can
be made on $\hat{N}^2$, without loosing or altering a single quanta of
modes A and B. It has been shown that the coupling between A and B as
in Hamiltonian \rf{ham1} displays several interesting features
\cite{inf}: (i) It leads to a complete states swapping (information
exchange) even at constant mean energy. (ii) If the states are
initially not entangled, the interaction will produce entanglement
only if one of the modes is prepared in a nonclassical state;
otherwise, if both modes are initially prepared as a direct product of
coherent states the interaction will not change this character in the
course of their evolution \cite{ent}.

In the interaction picture, Hamiltonian (\ref{1}) is written as
\be%
 H_I=\hbar\la\left(a^\dagger b +ab^\dagger \right)
+\hbar\chi(a^\dagger a+b^\dagger b)(c +c^\dagger) \, , \lb{ham2}
\ee%
where, resonance conditions, $\nu=\w _a-\w _b$ and $\nu'=\w _c$, must
be satisfied in order to eliminate the explicit time dependence
present in \rf{ham1}.

Next, suppose the monitor mode C (the meter in the terminology of
\cite{mi1}) is prepared in the vacuum state, so, during its evolution
it could absorb energy only from the intense classical field. The
fields A and B are considered, for the moment, to be in an arbitrary
state, that we write as an expansion in the number states basis $|m\rg
_A $ and $|n\rg _B$, $|\psi_{AB}(0)\rangle = \sum_{m,n} C_{m,n}
|m,n\rangle$ (from here on we omit the subscript $AB$). At time $t$
the evolved state of the AB system is represented by
\be%
|\Psi(t)\rangle = e^{-i\la t\left(a^\dagger b +ab^\dagger
\right)} \sum_{m,n }C_{m,n} |m,n \rg \ox e^{-i\chi t (m+n)(c
+c^\dagger)}|0\rg _C \, . \lb {psi1}
\ee%
Noting that $e^{-i\chi t (m+n)(c +c^\dagger)}|0\rg _{C} =|-i\chi t
(m+n)\rg _{C}$ is a coherent state restricted to the imaginary axis,
state \rf{psi1} can be written as
\be%
\label{2} |\Psi(t)\rangle=\sum_{m,n} C_{m,n} U_t |m,n \rg \ox |-i\chi
t (m+n)\rg   \, ,
\ee%
(here on we omit the subscript C) where $U_t = e^{-i\l
t\left(a^\dagger b +ab^\dagger \right)}$. In the next section we
discuss two forms of measurements on mode C.
%
\section{Monitoring AB fields trough measurement on C}
%
\subsection{Instantaneous projective measurement}
%
An ideal or instantaneous projective measurement (PM) of a system $
{\cal S}$ is associated to a chosen observable $\cal{A}$ (or a set of
commuting observables) of this system. If at time $t$ one of the
eigenvalues $A_m$ of $\cal{A}$ is realized, then the system state
$|\Psi (t)\rg $ is reduced to its corresponding eigenvector $|m\rg$.
The probability for that realization is $| \lg \Psi(t) |m\rg|^2$. This
is known as a measurement of the first kind, as  postulated by von
Neumann. In our model, if at time $t$ it is found that field C contains
exactly $k$ photons, the state is automatically projected on to the
number state $|k\rg $. As a consequence, the AB joint state `reduces'
instantaneously to the new state
\be%
\rho^{(k)}_{AB}(t)= \frac{\lg k| \rho(t) |k \rg}{{\rm Tr}_{AB}\lbk
\lg k|\rho(t)|k \rg \rbk}
\ee%
where $\rho(t)=|\Psi(t)\rg \langle\Psi(t)|$ and ${\rm Tr}_{AB}$
stands for the trace operation on the A and B fields operators. The
probability that at time $t$ the field C has $k$ quanta is $P(k,t)=$
Tr$_{AB} \lbk \lg k|\rho(t)|k \rg \rbk $. For the state given by
equation (\ref{2}) this probability is
\br%
P(k,t) &=& \frac{\lbk \chi t \rbk ^{2k}}{k!} \lg \Psi_{AB}(0) |
U^{\dag}_t \hat{N}^{2k}e^{-(\chi t)^2\hat{N}^2 }U_t |\Psi_{AB}(0)\rg
\nn
\\ &=& \frac{\lbk \chi t \rbk ^{2k}}{k!} \lg \Psi_{AB}(0) |
\hat{N}^{2k}e^{-(\chi t)^2\hat{N}^2 } |\Psi_{AB}(0)\rg =
\sum_{m,n=0}^{\infty}|C_{m,n}|^2\frac{1}{k!} \lpar \chi t(m+n) \rpar
^{2k}e^{-[\chi t(m+n)]^2} \, , \lb{probk1}
\er%
the second equality follows because $[U_t, \hat N] = 0$, so the counts
are independent of the evolution of AB modes, they are not affected by
the evolution of the AB system. The evolved state is reduced to
\be%
\rho_{AB}^{(k)}(t)=P^{-1}(k,t) \frac{1}{k!} \lpar \chi t\rpar
^{2k}\hat{N}^k e^{- \half (\chi t)^2 \hat{N}^2} U_t| \Psi_{AB}(0)\rg
\lg \Psi_{AB}(0)|U^{\dag}_t e^{- \half (\chi t)^2 \hat{N}^2}
\hat{N}^k  \, . \lb{rhoab1}
\ee%
or in terms of the statevector (excepting a phase factor),
\be%
|\Psi^{(k)}_{AB}(t) \rg = \frac{\lpar \chi t\rpar ^{k}}{\sqrt{k!
P(k,t)}}\hat{N}^k e^{- \half (\chi t)^2 \hat{N}^2} U_t|
\Psi_{AB}(0)\rg \, ,
\ee%
the purity of the state is maintained, however, entanglement is
created even if initially the state $|\Psi_{AB}(0)\rg$ is factorized.

The mean photon number of field C is closely related to the mean $\lg
{\hat N}^2 \rg$ of AB fields,
\be%
\ovl k = \sum_{k=0}^{\infty}k P(k,t) = (\chi t)^2
\sum_{m,n=0}^{\infty}|C_{m,n}|^2 (m+n)^2= (\chi t)^2 \lg {\hat N}^2
\rg \, ,\lb{kmedio0}
\ee%
so, a measurement on C allows to infer the mean squared QND variable
with a proportionality factor that goes as $t^2$. The variances are
related as
\be%
{\rm Var}(k) - \ovl k = (\chi t)^4 {\rm Var}({\hat N}^2) \, ,
\lb{Nvar}
\ee%
where ${\rm Var}(x)= \lg x^2 \rg - \lg x \rg ^2$.

If fields A and B are initially prepared in number states, $|
\Psi_{AB}(0)\rg = |m,n\rg $, the reduced state of the  AB field
becomes independent of $k$, evolving freely as
\be%
| \Psi_{AB}(t) \rg = U_t |m,n \rg \, \lb{psilivre}\, ,
\ee%
thus, not feeling at all the presence of field C; so, any
entanglement will only arise from the interaction between A and B and
not from a measurement on C. The probability for measuring $k$
photons in C at time $t$ will depend on $m$ and $n$ as $(m+n)^2$, the
probability distribution being Poissonian
\be%
P(k,t)=\frac{1}{k!} \lbk (\chi t)^2 (m+n)^2 \rbk ^k e^{-\lbk (\chi
t)^2 (m+n)^2 \rbk}\, .
\ee%
The mean photon number of C Eq. \rf{kmedio0} gives
\be%
\ovl k = (\chi t)^2  \lg {\hat N}^2 \rg = (\chi t)^2 (m+n)^2 \, ,
\lb{kmedio1}
\ee%
whereas
\be%
{\rm Var}(k) - \ovl k =(\chi t)^4 {\rm Var}({\hat N}^2) = 0 \, .
\ee%
Inversely, if one ignores in which states the AB modes were prepared,
but however verifies, through a measurement on C, that ${\rm Var}(k) =
\ovl k$, then one immediately infers that the AB system was prepared
in some number state $|m,n \rg $. If the modes A and B are prepared in
eigenstates of the QND variable the subsystem AB evolves independently
of C.

Now, if the initial states of the subsystems A and B are prepared as a
direct product of coherent states, $| \Psi_{AB}(0)\rg = |\alpha \rg
\ox |\beta \rg \equiv|\alpha , \beta \rg $, the unitary evolution does
not entangle the states of A and B, each one continues its evolution
as such,
\be%
|\Psi^{(k)}_{AB}(t) \rg = \frac{\lpar \chi t\rpar ^{k}}{\sqrt{k!
P(k,t)}}\hat{N}^k e^{- \half (\chi t)^2 \hat{N}^2} |\alpha (t), \beta
(t) \rg\label{pm1}
\ee%
where $U_t |\alpha,\beta \rangle = |\alf (t),\beta (t) \rg $, with
$\alpha (t)= \alpha \cos{\la t}-i\beta\sin{\la t}$ and $\beta (t)=
\beta \cos{\la t}-i\alpha\sin{\la t}$ (note that $|\alf (t)|^2 +|\beta
(t)|^2= |\alf |^2 +|\beta |^2 \equiv F$ is a constant of the motion).
Although the measurement on C affects the dynamics of the AB system by
entangling the states, this very ideal measurement does not allow
controlling the entanglement dynamics of the AB system, since the
outcome of the eigenvalue $k$ is probabilistic, according to the
distribution \rf{probk1}.

As in the previous case, the mean photon number of C will depend on
time as $t^2$,
\be%
\ovl k = (\chi t)^2 F \lpar F  + 1 \rpar \, , \lb{kmedio}
\ee%
the variance is given by
\be%
{\rm Var}(k) = (\chi t)^4 (4F^3 +6F^2 +F) \, . \lb{kvar}
\ee%
and $F$ can be obtained from the measurements on $\cal M$,
\be%
F=\lbk \lpar {\rm Var}(k)/(\chi t)^4 - 2 \ovl k/(\chi t)^2\rpar /
\lpar 4 \ovl k/(\chi t)^2 -1\rpar \rbk \, . \lb{F}
\ee%
where the right hand side, calculated from the experiment should be
time-independent.
%
\subsection{Generalized measurement by photocounting}
%
Experiments involving counting are not instantaneous and far from the
von Neumann idealization. For a realistic counting measurement, one
should consider that (i) For a given time interval $t$ the counting of
$k$ quanta occurs with some probability, or, it is not likely that
exactly $k$ quanta are counted in a predefined time $t$, actually,
there will be a distribution of time intervals. (ii) More importantly,
one has to consider that when one photon is counted the EM field will
have one photon less. The dynamics of this kind of process has to be
treated as a dissipative continuous measurement; this subject was well
addressed by Srinavas and Davies \cite{davies} and applied to several
situations in \cite{mi1,mi2,caves}. We follow closely these
references, describing the continuous photocount measurement in the
formalism of operations and effects \cite{kraus}.

The count of $k$ photons from the monitor mode in a time $t$ is
characterized by the linear operation $N_t(k)$, acting on the state of
the system,
\be%
\rho^{(k)}(t)=\frac{N_t(k)\rho (0)}{{\rm Tr}\lbk N_t(k)\rho(0) \rbk }
\ee%
where $\rho (0)$ is the state of the ABC system prior turning on the
counting process and $P(k,t)={\rm Tr} [N_t(k)\rho]$ is the
probability of counting $k$ photons in $t$. The linear operator
$N_t(k)$ is written in terms of two other operators, $S_t$ and $J$,
\be%
N_t(k)=\int_0^tdt_k\int_0^{t_k} dt_{k-1}\cdot\cdot\cdot\int_0^{t_2}
dt_1S_{t-t_k}JS_{t_k-t_{k-1}}\cdot\cdot\cdot JS_{t_1}\, ,
\ee%
where $S_t\equiv N_t(0)$ is a superoperator defined in terms of
ordinary Hilbert space operators $B_t$ as
\be
S_t\rho=B_t\rho B_t^\dagger.
\ee
$B_t$ is a semigroup element given in terms of the generator $Y$ as
$B_t=e^{Yt}$. As defined in \cite{davies} the generator is
\be%
Y=-\frac i\hbar H-R/2 \, ,
\ee%
where $H$ is the system Hamiltonian and $R$ is the rate operator
given by
\be%
R=\gamma c^\dagger c \, ,
\ee%
and the parameter $\gamma$ stands for the detector counting rate. The
theory becomes self contained by choosing $J$ as
\be%
J\rho=\gamma c\rho c^\dagger \, ,
\ee%
standing for the change of the field C due to loss of one counted
photon and $S_t$ is responsible for the state evolution between
counts.

In the interaction picture $Y$ becomes
\be%
Y=-i\l \left(a^\dagger b +ab^\dagger \right) -i\chi (a^\dagger
a+b^\dagger b)(c +c^\dagger) -\frac\gamma 2 c^\dagger c \, ,
\ee%
the first term contributes only to the free evolution of the AB
fields as a unitary evolution of the initial state. The monitor field
stands for the counting process, being present in the other terms,
thus the linear superoperator $N_t (k)$, acting on the initial state
$\sum_{m,n}C_{m,n} |m,n\rg  \ox |0\rg $, can be expressed as
\be%
N_t(k)={\cal U}\int_0^tdt_k\int_0^{t_k}
dt_{k-1}\cdot\cdot\cdot\int_0^{t_2}dt_1 \widetilde
S_{t-t_k}J\widetilde S_{t_k-t_{k-1}}\cdot\cdot\cdot J\widetilde
S_{t_1},
\ee%
where
\be%
\widetilde S_t\bullet=e^{\widetilde Y_1 t}\bullet e^{\widetilde
Y_2^\dagger t}
\ee%
with
\begin{eqnarray}%
\widetilde Y_1&=& -i\chi (m+n)(c +c^\dagger) -\frac\gamma 2 c^\dagger
c\\ \widetilde Y_2&=& -i\chi (m'+n')(c +c^\dagger) -\frac\gamma 2
c^\dagger c.
\end{eqnarray}%
We have defined ${\cal U} _t$ as the superoperator for the coherent
evolution of modes A and B
\be%
{\cal U}_t\bullet= e^{-i\la t(a^\dagger b+ab^\dagger)}\bullet e^{i\la t
(a^\dagger b+ab^\dagger)}.
\ee%
After doing some algebraic manipulations we find that for $k$ counts
on C the state for the ABC system becomes
\br%
\rho^{(k)}(t) &=& \frac{1}{P(k,t)} \frac{\lbk 2 g(t) \rbk ^k}{k!}
\sum_{m,n,m',n'}C_{m,n}C^*_{m',n'}(m+n)^k (m'+n')^k  \exp [ A(t)+
A'(t) \nn \nonumber\\
&& + \half |z(t)|^2 + \half |z'(t)|^2 ]  {\cal U}_t \lbk| m,n\rangle
\langle m',n'| \rbk \lbk | z(t) \rangle \langle z'(t)| \rbk  \, ,
\lb{post}
\er%
where $g(t) = (2 \chi ^2 / \g ^2) \lpar -3 + \g t + 4e^{- \g t /2} -
e^{- \g t} \rpar $, $ z(t)=  (-2i\chi /\g) (m+n)(1-e^{-\g t/2}) $ is
the label of the coherent state and $A(t) =(-2\chi^2/\g ^2)(m+n)^2
\lbk \g t- 2(1-e^{\g t/2}) \rbk$. $z'(t)$ and $A'(t)$ stand for the
same expressions but with $m'+n'$ instead of $m+n$. The normalization
function $P(k,t)$ is the probability for $k$ counted photons in time
$t$,
\be%
P(k,t) = \frac{\lbk 2 g(t) \rbk ^k}{k!} \lg \Psi_{AB}(0) |
\hat{N}^{2k}e^{-2g(t)\hat{N}^2 } |\Psi_{AB}(0)\rg  \, \lb{probk2},
\ee%
the same as in the ideal PM, Eq. \rf{probk1}, however the factor
$(\chi t)^2 $ is now replaced by $ 2 g(t)$. An important difference
arises here, for $\g t \gg 1$ we have $2g(t) \simeq (2\chi/ \g )^2 \g
t$, the time appearing linearly while in the ideal PM it is quadratic.
The expressions for the mean and variance of $k$ are as \rf{kmedio}
and \rf{kvar} however with $ (2\chi/ \g )^2 \g t$ substituting $(\chi
t)^2$.

The mixed state $\rho^{(k)}(t)$ is a post-measurement state for
counting $k$ photons. The pre-measurement state is obtained by summing
over all the possible outcomes, $\sum_k N_t (k) \rho(0)$ \cite{caves}.

The state of the AB system is obtained by tracing over the monitor
mode,
\begin{eqnarray}%
\rho _{AB}^{(k)}(t)&=&\frac{1}{P(k,t)} \frac{\lbk 2 g(t)\rbk ^k
}{k!}\hat{N}^k
e^{-h(t)\hat{N}^2} \nn \\%
&& \x \lbr e^{\frac{4 \chi ^2 }{\g ^2}(1- e^{-\g t /2 })^2 \hat{N}
\bullet \hat{N}} \lbk {\cal U} _t \lpar  |\Psi_{AB}(0)\rangle \langle
\Psi_{AB}(0)| \rpar \rbk \rbr e^{-h(t) \hat{N}^2} \hat{N}^k \, , \lb
{rhoab2}
\end{eqnarray}%
where $h(t)=(2 \chi ^2 / \g ^2)[\g t -2(1- e^{-\g t/2 })] $, and
\be%
e^{\frac{4 \chi ^2 }{\g ^2}(1- e^{-\g t /2 })^2 \hat{N} \bullet
\hat{N}} \hat O \equiv \sum_{l=0}^{\infty}\frac{\lbk \frac{4 \chi ^2
}{\g ^2}(1- e^{-\g t /2 })^2 \rbk ^l}{l!}\hat{N}^l \hat O \hat{N}^l
\, .
\ee%

Preparing the modes A and B in number states the mode C has its
dynamical evolution decoupled from the AB system. As in the ideal PM,
Eq. (\ref{psilivre}), the AB modes states evolve freely, not being
affected at all by the counting process (due to the QND variable).

For the modes A and B prepared in coherent states, $ |\Psi_{AB}(0) \rg
=|\alpha , \beta \rangle$, the unitary evolution does not change this
character, each state continues to evolve as such ($|\alf (t),\beta
(t)\rangle \langle \alf (t),\beta (t)|$), however, the photocount on C
affects the dynamics introducing a another effect: besides the
entanglement, as in the ideal PM, the factor in brackets in
\rf{rhoab2} mixes the states. The number of counted photons in mode C
determines the selection of a specific entangled state of the AB field
and the degree of its entanglement. This will depend on $\chi$, the
counting rate of the detector $\g$, the number of counts $k$, the time
$t$, and there will also be a dependence on the initial state through
the total number of photons operator ${\hat N}$. So, controlling these
quantities entails the correlated state \rf{rhoab2}. It is worth
stressing that time interval for counting exactly $k$ photons is
probabilistic and $P(k,t)$ is its (non-normalized) distribution
function.
%
\subsection{Degree of entanglement and mixing}
%
It is well known that a precise measure of the degree of entanglement
is not available for continuous variables mixed states \cite{cirac}.
We saw that a straightforward application of the photocount
measurement process, may be used for producing and determining the
degree of a $k$-entangled state of AB modes. Formally, for the AB
system, prepared initially in coherent states, the dissipative
(nonunitary) character of the evolution is induced by the counting
process,  being responsible for the interplay between the
entanglement, mixing and decoherence in \rf{rhoab2}.

For the lowest values of $k$ the probability distributions $P(k,t)$
are depicted in Fig. 1 for $|\alpha|^2=|\beta|^2=5$. One perceives
that for $ k=0$ the highest values of the probability occur for
$\gamma t\ll 1$, however for $k\neq 0$ the probabilities attain
maximum values at different times $t_m$. For $\gamma t\ll 1$ one has
$h(t) \approx (\chi t)^2 /2 $, $4(\chi /\g)^2(1 - e^{-\g t /2})^2
\approx (\chi t)^2$ and $2g(t) \approx (\chi /\g)^2 (\g t)^3 /3 $, so
state (\ref{rhoab2}) does not mix substantially for small time
intervals, it can be written as
\be%
|\Psi^{(k)}_{AB}(t) \rg \approx P^{-1/2}(k,t)\hat{N}^k|\alpha
(t),\beta (t) \ra \, , \lb{psiab4}
\ee%
(here, $P(k,t)= \lg \alpha (t),\beta (t) |\hat{N}^{2k} |\alpha
(t),\beta (t) \rg $) and, excepting for $k=0$ ($ P(k,0)= \delta_
{k,0}$), the counting process generates entanglement.

Comparing state (\ref{psiab4}) with (\ref{pm1}) for $ (\chi /\gamma)
\gamma t\ll 1$, we see that they are very likely, thus for small times
they cannot be distinguished and the entanglement will be stronger the
higher the number of counted photons. We give a quantitative picture
of entanglement by calculating the excess entropy \cite{lindblad}, defined as
\be%
I=S_A+S_B-S_{AB},%
\ee%
where $S_A$ ($S_B$) is the entropy of the mode A (B)(associated to
$\rho^{k}_A (t) = {\rm Tr}_B \rho^{(k)}_{AB}(t) $) and $S_{AB}$ is the
entropy of the AB system, where $S \equiv 1 - {\rm Tr} \rho ^2 $.
The excess entropy measures the information contained in the correlation between modes A and B.
Its lower and upper bounds are obtained by the Araki-Lieb \cite{araki} inequality
\be
|S_A-S_B|\le S_{AB}\le S_A+S_B,
\ee
namely,
\be\label{ineq1}
0\le I\le 2\; \text{min}(S_A,S_B).
\ee
Remark that $I$ measures the correlation between A and B, without resolving between
classical or quantum 
correlation (entanglement).
However, when the joint system is in a pure state $S_{AB}=0$ and $S_A=S_B$, 
the inequality (\ref{ineq1}) reduces to
\be
0\le I\le 2 S_A.
\ee
and any correlation given by
 $I>0$ will be due to entanglement. Thus, the measure of $I$ together with the system degree
of purity allows one to distinguish between classes of entangled states, even though it does
not give a precise borderline for separability of mixed entangled states.

 The
excess entropy for the system here considered is plotted in Fig. 2 as function of $k$ 
for most probable time of a $k$-event (crosses for
\rf{rhoab2}) and for initial times ($\gamma t\ll 1$) (filled squares for \rf{psiab4}).
 It is verified that
the counts correlate the subsystems more intensely the higher is the
number of counted photons. Since state (\ref{psiab4}) is pure
($S_{AB}=0 $), this correlation characterizes a maximal entanglement. When detections occur at initial
time the state of the joint system AB is left in a pure entangled state and the degree of entanglement is directly
proportional to the number of counted photons.
The excess entropy for state \rf{rhoab2} is calculated at times
$t^{(k)}_m$ (for which $P(k,t^{(k)}_m) $ is maximum - as in Fig. 1) and
one sees that the correlation is stronger than for state \rf{psiab4}.
However, as state (\ref{rhoab2}) is mixed one also need to measure its purity ($S_{AB}$), as
depicted in Fig. 3, for several values of $k$ and at times $t^{(k)}_m
$. The combined results of Figs. 2 and 3 for the mixed state entanglement, 
show that the higher the
number of counted photons, the more correlated becomes the AB state,
and for $k \gg 1$ one has $I \approx 1$. However, the states more
likely to occur at times $t^{(k)}_m$  are not pure anymore and for
high $k$'s, characterized by $I\lesssim1$ and $S_{AB} \lesssim1$,
the state \rf{rhoab2} becomes classically correlated (separable),
while for $0<S_{AB}<1$ it is non-separable, which characterizes
non-maximal entanglement.


How this process happens can be represented by the following example, which exactly matches our results for
both the excess entropy and purity.
Considering the condition of separability,
\be%
\rho_{AB}=\sum_i p_i \sig ^{(i)} _A \ox \sig ^{(i)} _B\, , \qquad
\sum_i p_i =1 \, , \lb{rhoab5}
\ee%
with $\rho_{AB}$ acting on ${\cal H}_A \otimes{\cal H}_B$; $\dim {\cal
H}_A= N_A$ and $\dim{\cal H}_B = N_B$,
then,
\be%
I=1- \sum_{i,j} p_i p_j \lbk {\rm Tr}_A \sig ^{(i)} _A \sig ^{(j)} _A +
{\rm Tr}_B \sig ^{(i)} _B \sig ^{(j)} _B  - \lpar {\rm Tr}_A \sig
^{(i)} _A \sig ^{(j)} _A \rpar \lpar {\rm Tr}_B \sig ^{(i)} _B \sig
^{(j)} _B \rpar \rbk \, .
\ee%
When
\be%
\sig ^{(i)} _A = \frac {\hat {1}_A}{N_A} \qquad {\rm and} \qquad \sig
^{(i)} _B = \frac {\hat{1}_B}{N_B} \lb{sigab}
\ee%
one obtains
\be%
I=1- \lpar N_A ^{-1}+ N_B ^{-1}- \lpar N_A  N_B \rpar ^{-1} \rpar
\qquad {\rm and} \qquad  S_{AB} = 1 - \lpar N_A  N_B \rpar ^{-1}
\ee%
For $N_A , N_B \rightarrow\infty$ (as it is for coherent states) both
quantities go to $1$. So, for a $k$ sufficiently large the state
\rf{rhoab2} becomes separable. State (\ref{rhoab5}) (together with
\rf{sigab}) corresponds to an equiprobable ensemble of states.

In that way, we saw that a full range of states may be generated, with
different degree of entanglement, from maximally entangled to
separable states. So, the protocol for producing a specific state
\rf{rhoab2} consists in turning off the interaction when an  a priori
selected value $k$ is attained at a time $\ovl t$.

Besides, the photocount measurement theory and the specific Hamiltonian
\rf{ham2} also allow to extract information about the AB system by
counting the photons of C. Due to the nature of the coupling between
the monitor and the AB system, the counting on C gives the maximal
information about any function of $\hat N^2$. Then it is immediate to
put the question:  How much information may the photocount
distribution give about the initial quantum states $|\Psi_{AB}(0) \rg
$? This is discussed bellow.
%
\section{Probing light with light}
%
Here we look at the counting problem as a spectroscopic measure of
the state of the AB system. Light state spectroscopy, or light
probing via atomic deflection was discussed by M. Freyberger and A.
M. Herkommer \cite{herkommer}. Probability distribution of
transversal momenta of two-level atoms deflected by an EM field in a
cavity allows the complete knowledge of the field quantum state. Is
it possible to use a similar strategy with fields only, but now
instead of the beam deflection, the photocount playing the role
revealing the AB state?

To make this point clear let us consider the average counted photon
number
\be%
\ovl k=u(t)\lg{\hat N}^2 \rg =u(t) \sum_{m,n} |C_{m,n}|^2 (m+n)^2
\ee%
where $ u(t)=\left(2 \chi / \gamma\right)^2 (-e^{-\gamma
t}+4e^{-\gamma t/2}+\gamma t-3) $. For $\gamma t\gg 1$, $u(t)=\left(2
\chi / \gamma\right)^2 \g t$ and we can write
\be%
\ovl k = \lbk \lpar 2\chi /\g \rpar^2 \g t \rbk \sum_{m,n} |C_{m,n}|^2
(m+n)^2 \lb{kmedio2}
\ee%
and $\lg \hat{N}^2 \rg$ can be inferred directly by the mean counted
photons, this being more precise the larger $\gamma t$ is because one
gets a linear relation between time $t$ and $\ovl k$. This point was
actually pointed by Milburn and Walls \cite{mi1} for a single mode
coupled to C. Note that differently from the projective measurement
where $\ovl k$ increases with $t^2$, Eq. \rf{kmedio1}, in the
non-ideal measurement $\ovl k$ increases linearly with time. We can
build up higher-order moments of $k$ in a similar fashion, obtaining,
\be%
\ovl{k^r}=\sum_{m,n}|C_{m,n}|^2 e^{-u(t)(m+n)^2}\left( u \frac{d}{d
u}\right)^r e^{u(t)(m+n)^2} \, ,
\ee%
or keeping only the higher power in $\lbk \lpar 2\chi / \g \rpar ^2
\lpar \g t \rpar  \rbk\gg 1$ for conveniently chosen time $t$, we have
\be%
\ovl{k^r} = \lbk \lpar 2\chi / \g \rpar ^2  \lpar \g t \rpar  \rbk ^r
\sum_{m,n} |C_{m,n}|^2 (m+n)^{2r} .
\ee%
Calling
\be%
\kappa=k/\lbk \lpar 2\chi / \g \rpar ^2  \lpar \g t \rpar \rbk
\lb{kappa} \, ,
\ee%
multiplying both sides of the above equation by $(-1)^r x^{2r}/r!$
($0 \leq x < 2 \pi$) and summing over $r$ we get a new function,
which a Fourier expansion of the squared moduli coefficients,
\begin{equation}%
H(x)=\sum_{r=0}^{\infty} \frac{(-1)^r x^{2r}\ovl{\kappa^r}
}{r!}=\sum_{m,n}|C_{m,n}|^2 \cos{x(m+n)},
\end{equation}%
which is always bounded, with values in the interval $[-1,1]$ since
$\sum_{m,n}|C_{m,n}|^2 = 1$. The left hand side is calculated from
the experimental photocounts, thus a specific relation between a
particular set of the $|C_{m,n}|^2$ may be inferred in certain cases,
as to be see below. Doing an inverse  Fourier transform we get
\be
{\cal C}(j)=\frac{1}{\pi}\int_0^{2\pi} \cos{jx} H(x) dx =
\sum_{m=0}^j |C_{m,j-m}|^2 \, .
\ee%
which is the only information on the initial state of the AB system
one gets by photocounting on C. If the fields A and B are initially
disentangled, $|C_{m,n}|^2= |A_{m}|^2|B_{n}|^2$, then in order to
determine the coefficients of, for instance, the initial state of
mode A, $|A_{m}|^2$, one should set the initial state of B in the
vacuum state (or any other number state), since $|B_{0}|^2=1$ and all
other coefficient being zero, so one obtains
\be%
|A_{m}|^2={\cal C}(m)\,.
\ee%
This strategy has its limitation since only the moduli of the
coefficients can be obtained, being the maximum information about the
initial state of field A that one can obtain by counting on C.

Particularly interesting states to be addressed are the initially
entangled states.

(a) We first consider the case when both modes are prepared in the
superposition of perfect anti-correlation
\be%
|\Psi_{AB}\ra=\sum_{n=0}^N C_{N-n,n} |N-n,n\ra \, , \lb{psiNn}
\ee%
defined as a limited sum of the photon number state, and
$\sum_{n=0}^N |C_{N-n,n}|^2=1$.
One illustrative example of this kind of
state is for entangled qubits (${\rm dim} [{\cal H_A \otimes
H_B}]=2\otimes 2$), $|\Psi_{AB}\rangle= C_{1,0} |1,0\ra+C_{1,0}|0,1\ra$.

For state \rf{psiNn} the probability of counting $k$ photons is
independent of the coefficients $C_{N-n,n}$,
\begin{eqnarray}%
P(k,t)&=&\frac{N^{2k}}{k!}\left(\frac{2\chi}{\gamma}\right)^{2k}(-e^{-\gamma
t}+4e^{-\gamma t/2}+\gamma t-3)^k\nonumber\\ &&
\times\exp{\left[-\left(2\chi /\gamma\right)^2 N^2 (-e^{-\gamma
t}+4e^{-\gamma t/2}+\gamma t-3)\right]}\, ,
\end{eqnarray}%
so, all higher moments are determined from the first one, the mean
counted photons is a precise measurement since the variance is zero
and consequently the squared coefficients $|C_{N-n,n}|^2 $ cannot be
determined. For $\g t \gg 1$ the expression for the $r$-moments
(expressed as $\kappa$, see \rf{kappa}) of counted photons will be
\be%
\ovl{\kappa^r}=N^{2r} = \lpar \ovl{\kappa} \rpar ^r \, .
\ee%

(b)  Now, let us consider the case when both modes are prepared in
the superposition of perfect correlation
\be%
|\Psi_{AB}\ra= \sum_{n=0}^{N/2} C_{n,n} |n,n\ra \, . \lb{psinn}
\ee%
Again, for qubits, $|\Psi_{AB} \ra=  C_{0,0} |0,0\ra+C_{1,1}|1,1\ra$.
For the state \rf{psinn} the probability of counting $k$ photons is
\begin{eqnarray}%
P(k,t)&=&\frac{1}{k!}\left(\frac{2\chi}{\gamma}\right)^{2k}
\sum_{n=0}^{N/2}|C_{n,n}|^2(2n)^{2k} (-e^{-\gamma t}+4e^{-\gamma
t/2}+\gamma t-3)^k\nonumber\\ &&\times
\exp{\left[\left(\frac{2\chi}{\gamma}\right)^2(2n)^2 (-e^{-\gamma
t}+4e^{-\gamma t/2}+\gamma t-3)\right]}
\end{eqnarray}%
and for $\g t \gg 1$ the moments of counted photons (expressed in
terms of $\kappa$) will be
\be%
\ovl{\kappa^r}=\sum_{n=0}^{N/2} |C_{n,n}|^2(2n)^{2r}\label{coef}
\ee%
and all coefficients squared moduli can be determined,
\be%
{\cal C}(2n)=|C_{n,n}|^2\, , \qquad n=0,1,2,...,N/2 \, .
\ee%
When $N\rightarrow \infty$ and
\be%
C_{n,n}= \frac{\lpar \tanh r \rpar^n}{\cosh r} \, ,
\ee%
state \rf{psinn} is a two-mode squeezed state ($r$ is the squeezing
parameter) used to establish the quantum channel for the continuous
variable teleportation, as reported in \cite{furusawa}. So, all terms
can be precisely determined since all depend on the parameter $r$ that
can be inferred from the relations between the coefficients.

We call attention to the fact that although this procedure permits to
determine the coefficient moduli $|C_{m,m}|$ of state \rf{psinn}, it
does not allow to distinguish this state from the mixed state
\be%
\rho_{AB}(0)= \sum_{n=0}^{N/2}|C_{n,n}|^2 |n,n\ra \langle n,n|
\lb{convex}
\ee%
because the same outcomes are inferred, Eq. (\ref{coef}), for the
$r$-moments.
%
\section{Summary and concluding remarks}
%
We have considered a monitor subsystem ${\cal M}$ (an EM mode, C)
coupled to a system of interest ${\cal S}$, (two interacting modes, A
and B), where the interacting term in the hamiltonian contains a QND
variable of ${\cal S}$: the total quanta of modes A and B. We proposed
a nondeterministic entanglement generation protocol of the two modes,
A and B, based on the continuous photodetection theory. By counting
destructively $k$ photons of the mode C the amount of entanglement of
the joint state of modes AB, prepared initially in coherent states,
can be controlled. Due to the dissipative character of the non-ideal
photodectection model, nonmaximally entangled states (mixtures) are
generated. The distribution function for counting photons $k$ and time
intervals $t$ is well defined allowing the determination of the most
probable time for the occurrence for each value of $k$.

For the sake of comparison we also investigated the case of an ideal or
projective measurement (instantaneous) as discussed by von Neumann.
The projection of the ${\cal S-M}$ state in a photon number operator
eigenstate $|k\rg$ pure entangled of A and B, however no control can
be done, because the realization of eigenstate cannot be fixed a
priori, it occurs probabilistically.

The post-selected counting distribution function allows calculating the
moments of the counted photons of C, which, for $\gamma t\gg 1$ are
closely related to the moments of the squared number of photons of both
modes, ${\hat N}^2$, which is a constant of the motion. So, also higher
moments of ${\hat N}^2$ can be inferred in a nondemolition
measurement. This has a immediate use, as probing the state of the AB
system by means of the counting distribution function. Examples were
given, and as expected, the extracted information showed to be useful
for partial or total reconstruction of the initial state of modes A
and B. Since only the squared modulus of the coefficients of the state
are given, we remarked that such procedure is not able to distinguish
systems between pure or mixed states prepared as maximally entangled.

Although the system here discussed is constituted of EM field modes,
the couplings may possibly be realized in other systems such as
vibrational degrees of freedom of trapped ions \cite{knight} or even
Bose-Einstein condensates \cite{condensate}. In both cases the
coupling of the atomic system with a light field (monitor mode) is
able to entangle atomic systems. It is also possible to probe the
atomic system state by photocounting on a light beam. In such cases we
expect collisions to be important if not restrictive to the method. An
obvious extension of the protocol here proposed is to use the
classical information achieved to control the AB system  state in a
continuous feedback process \cite{wiseman}. This allows the coherent
control of state entanglement of systems, and would be, indeed, useful for
quantum information processing \cite{qinf}.
%
\acknowledgments{This work was supported by FAPESP under contract $\#$
00/15084-5. MCO acknowledges FAPESP S\~ao Paulo) for total financial
support; LFS acknowledges total financial support from CAPES
(Bras\'{\i}lia) and SSM acknowledges partial financial support from
CNPq (Bras\'{\i}lia)}.

\newpage
{\bf figures}






\vspace{-0.7cm} 
\begin{figure}
\centerline{$\;$\hskip 0 truecm\psfig{figure=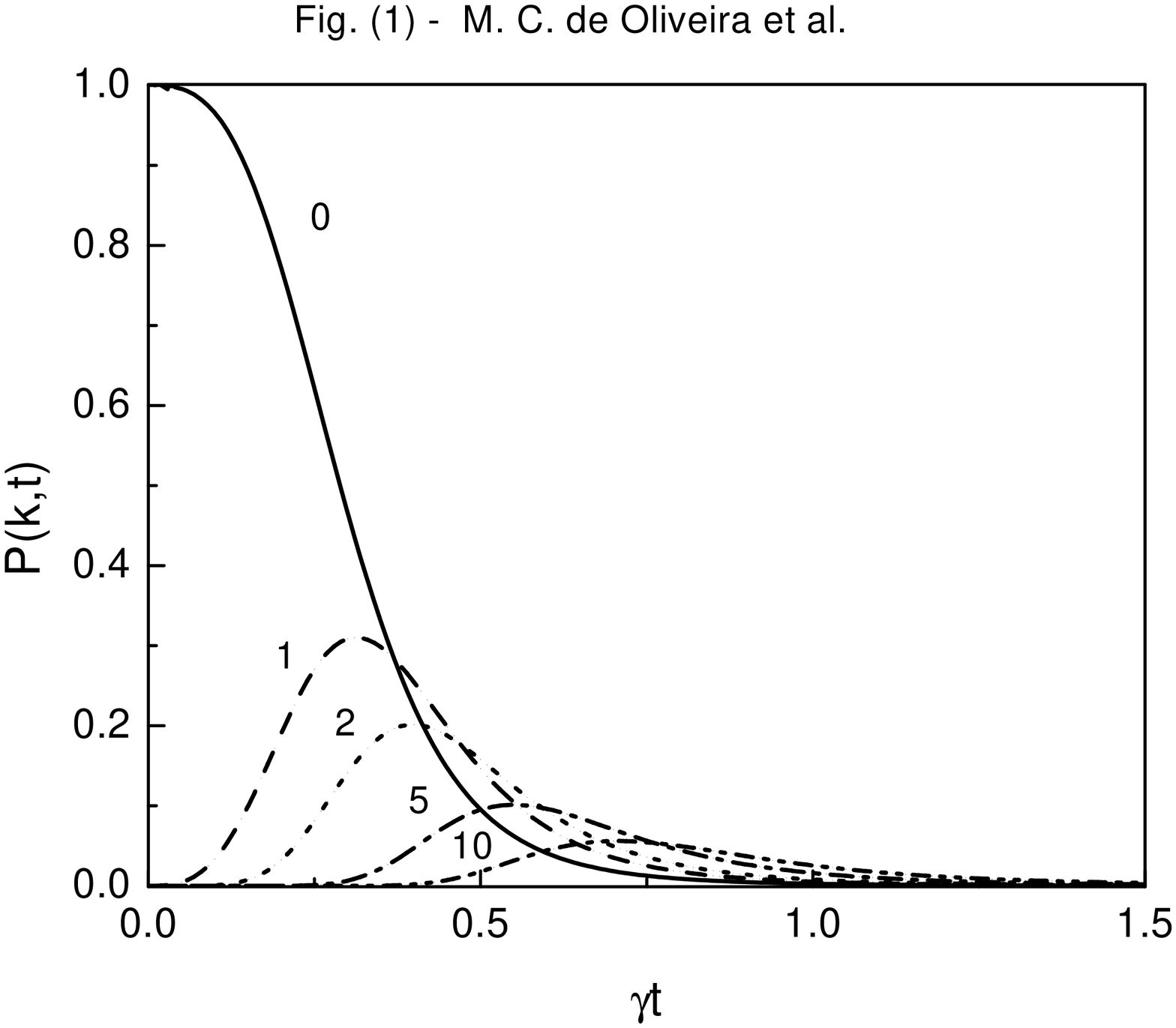,height=6.5cm,angle=0}}
\centerline{\caption{Photocount probability distribution for initial coherent
states for modes A and B, with $|\alpha|^2=|\beta|^2=5$. Numbers above
the curves indicate the counted photons. $\gamma t$ is a dimensionless time scale.
}}
\label{fig1}
\end{figure}

\vspace{-0.7cm} 
\begin{figure}
\centerline{$\;$\hskip 0 truecm\psfig{figure=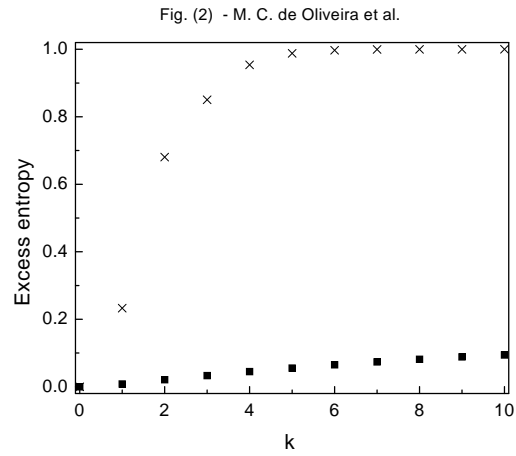,height=6.5cm,angle=0}}
\centerline{\caption{The excess entropy at the short time limit (filled
squares) and excess entropy at most probable time of the $k$-event
(crosses). For $k=0,...10$, at $\gamma t= 0.0$,
$0.32$, $0.4$, $0.46$, $0.51$, $0.55$, $0.59$, $0.62$, $0.65$, $0.68$, $0.71$.
}}
\label{fig1}
\end{figure}
\vspace{-0.7cm} 
\begin{figure}
\centerline{$\;$\hskip 0 truecm\psfig{figure=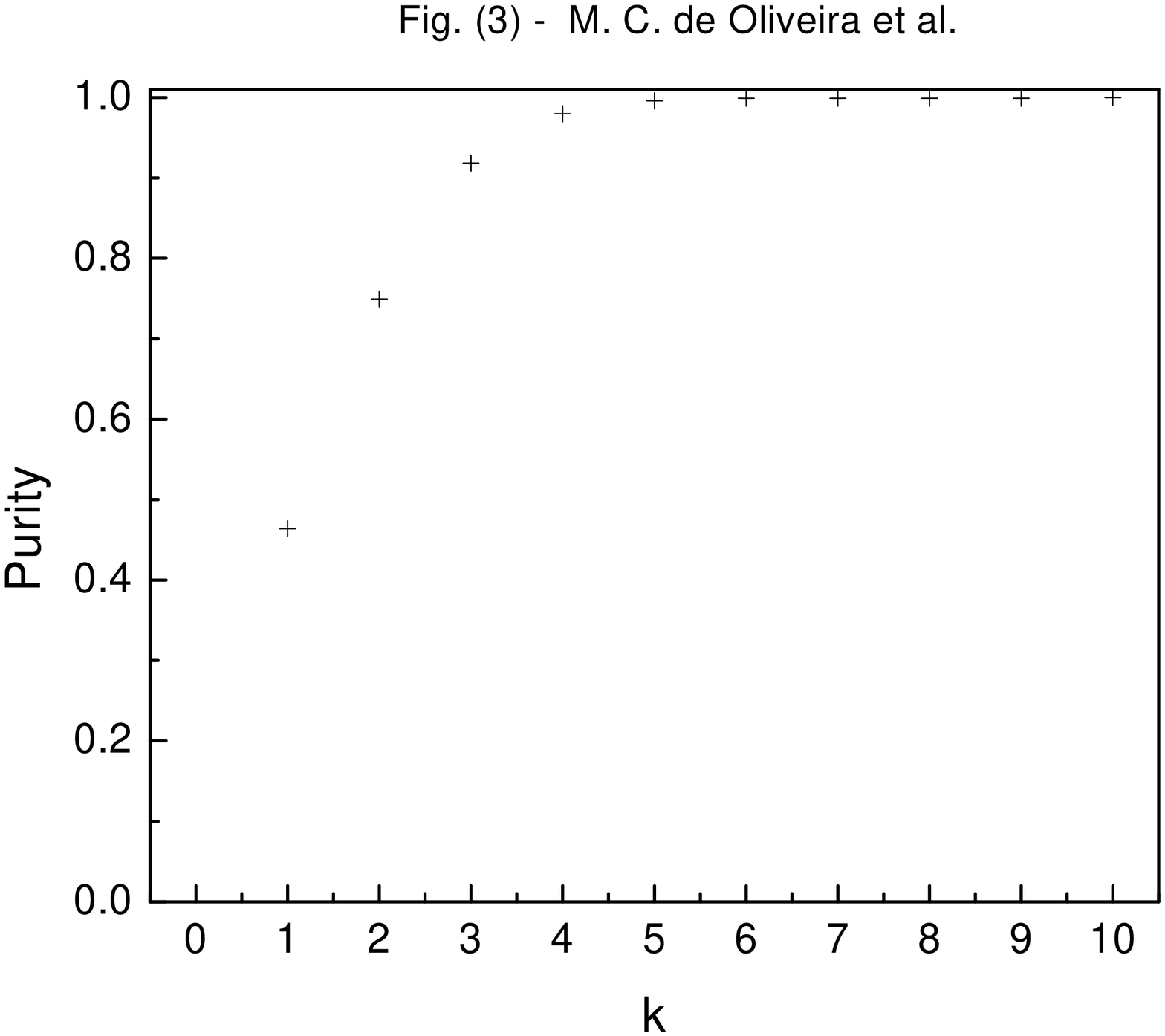,height=6.5 cm,angle=0}}
\centerline{\caption{Measure of purity of the $k$-event, at same times as in
Fig. 2.}}
\label{fig1}
\end{figure}


\begin{references}
%
\bibitem{qinf} M.A.  Nielsen and I.L. Chuang  {\it Quantum Computation and
Quantum Information} (Cambridge University Press, UK,2000).
\bibitem{ent} M.C. de Oliveira and W.J. Munro {\it Quantum resources and information
exchange in deterministic entanglement formation}, submitted for publication.
\bibitem{cirac} M. Lewenstein, B. Kraus, P. Horodecki, and J.I. Cirac, Phys. Rev. A {\bf 63}, 044304 (2001)
\bibitem{kwiat} P.G. Kwiat, K. Mattle, H. Weinfurter, A. Zeilinger,
A.V. Sergienko, and Y. Shih, Phys. Rev. Lett. {\bf 75}, 4337 (1995).
\bibitem{white} A.G. White, D.F.V. James, P.H. Eberhard, and P.G. Kwiat,
Phys. Rev. Lett. {\bf 83}, 3103 (1999).
\bibitem{hagley}E. Hagley, X. Ma\^\i tre, G. Nogues, C. Wunderlich, M. Brune, J. M. Raimond, 
and S. Haroche, Phys. Rev. Lett. {\bf 79}, 1 (1997).
\bibitem{rauschenbeutel} A. Rauschenbeutel, G. Nogues, S. Osnaghi, P. Bertet, M. Brune,
J.M. Raimond, and S. Haroche, Science {\bf 288}, 2024 (2000).
\bibitem{sackett} C.A. Sackett, D. Kielpinski, B.E. King, C. Langer, V. Meyer, C.J. Myatt,
M. Rowe, Q.A. Turchette, W.M. Itano, D.J. Wineland, and C. Monroe, Nature {\bf 404}, 256 (2000).
\bibitem{nature} B. Julsgaard, A. Koxhekin, and E.S. Polzik, Nature {\bf 413}, 400 (2001).
\bibitem{davies} M.D. Srinavas and E.B. Davies, Opt. Acta {\bf28}, 981 (1981).
\bibitem{yariv} A. Yariv, {\it Quantum Electronics}, (John Wiley \& Sons, USA, 1989).
\bibitem{kimble}Q.A.~Turchette, C.J.~Hood,
 W.~Lange, H.~Mabuchi, and H.J.~Kimble,
Phys. Rev. Lett. {\bf 75}, 4710 (1995).
\bibitem{parkins} S. Rebic, S. M. Tan, A. S. Parkins, and D.F. Walls, J. Opt. B {\bf 1}, 490 (1999).
\bibitem{milburnli} D. F. Walls and G. J. Milburn, {\it Quantum Optics},
(Springer-Verlag, Berlin, 1995).
\bibitem{scully} M. O. Scully and M. S. Zubairy {\it Quantum Optics}, (Cambridge University Press, UK) 1997
\bibitem{inf} M.C. de Oliveira, V.V. Dodonov and S.S. Mizrahi, J. Opt. B {\bf 1}, 610 (1999).
\bibitem{mi1} G.J. Milburn and D.F. Walls, Phys. Rev. A {\bf 30}, 56 (1984).
\bibitem{mi2} C.A. Holmes, G.J. Milburn, and D.F. Walls, Phys. Rev. A {\bf 39}, 2493 (1989).
\bibitem{caves} C.M. Caves and G.J. Milburn, Phys. Rev. A {\bf 36}, 5543 (1987).
\bibitem{kraus} K. Kraus, {\it States, Effects, and Operations} (Springer-Verlag, Berlin, 1984).
\bibitem{lindblad}G. Lindblad, {\it Non-equilibrium entropy and irreversibility}, (Reidel,Dordrecht,1983).
\bibitem{araki} H. Araki and E.H. Lieb, Commun. Math. Phys. {\bf18}, 160 (1970).
\bibitem{herkommer} M. Freyberger and A.M. Herkommer, Phys. Rev. Lett. {\bf 72}, 1952 (1994).
\bibitem{furusawa} A. Furusawa, J.L. S\o rensen, S.L. Braunstein, C.A.
Fuchs, H.J. Kimble and E.S. Polzik, Science {\bf 282}, 706 (1998).
\bibitem{knight} J. Steinbach, J. Twamley, and P.L. Knight, Phys. Rev. A {\bf 56}, 4815 (1997).
\bibitem{condensate} A.S. Parkins and D.F. Walls, Phys. Rep. {\bf 303}, 1 (1998).
\bibitem{wiseman} H.M. Wiseman, {\it Quantum Trajectories and Feedback}(Phd Thesis, 1994).
\end{references}
\end{document}